\documentclass[aps, preprint, superscriptaddress,floatfix, nofootinbib]{revtex4}

\usepackage{epsf}
\usepackage{graphicx}
\usepackage{amsmath}

\def\muegamma{\mu \rightarrow e + \gamma}

\def\beq{\begin{equation}}
\def\eeq{\end{equation}}
\def\beqa{\begin{eqnarray}}
\def\eeqa{\end{eqnarray}}

\begin{document}

\title{Yukawa Textures, Neutrino Masses and Horava-Witten M-Theory}

\author{R. Arnowitt}
\email{arnowitt@physics.tamu.edu} \affiliation{Center For
Theoretical Physics,  Department of Physics, \\Texas A$\&$M
University, College Station, TX, 77843-4242, USA}

\author{B. Dutta}
\email{duttabh@yogi.phys.uregina.ca} \affiliation{Department of
Physics, University of Regina, \\Regina SK, S4S 0A2, Canada}

\author{B. Hu}
\email{b-hu@physics.tamu.edu} \affiliation{Center For Theoretical
Physics,  Department of Physics, \\Texas A$\&$M University,
College Station, TX, 77843-4242, USA}

\date{\today}

\begin{abstract}
We consider the Horava-Witten based model with 5-branes situated
near the distant orbifold plane and with vanishing instanton
numbers on the physical plane. This model has a toric fibered
Calabi-Yau with del Pezzo base $dP_7$ which allows three
generations with Standard Model gauge group at the GUT scale.
Previous analysis showed that the quark hierarchy at the
electroweak scale could be achieved qualitatively without undue
fine tuning due to the effects of the 5-branes on the Kahler
potential. We extend here this analysis to include the leptons. A
new mechanism is introduced to obtain neutrino masses by assuming
massless right handed neutrinos exist in the particle spectrum,
which allows a cubic holomorphic term to exist in the Kahler
metric, $l_L H_2 \nu_R$, scaled by the $11D$ Planck mass. After
transferring this term to the superpotential, this term gives rise
to neutrino masses of the correct size at the electroweak scale.
With natural choices of  the Yukawa and Kahler matrix entries, it
is possible to fit all mass, CKM and MNS experimental data. The
model predicts $\mu \rightarrow e + \gamma$ decay at a rate that
should be detectable for much of the SUSY parameter space in the
next round of experiments.
\end{abstract}

\maketitle

\section{Introduction}

While the Standard Model (SM) has been successful in fitting all
current accelerator data, the origin  of the quark and lepton mass
spectrum remains a puzzle requiring further understanding. Thus
the explanation of the striking hierarchy of masses (e. g. the up
to top quark mass ratio is $m_u/m_t\simeq 10^{-5}$) and the
hierarchy of elements in the CKM matrix all are beyond the scope
of the Standard Model. The matter has been further exacerbated by
the discovery of neutrino masses, since now in addition there is
need for an explanation of the MNS matrix as well as the origin of
the very tiny neutrino masses. A large number of  suggestions
exist in the literature attempting to explain these properties of
quarks and leptons. One approach, starting perhaps with the work
of Georgi and Jarlskog \cite{gj}, suggests that the fundamental
origin of quark and lepton masses is to be found at high energies,
i. e. the GUT scale, $M_G \cong 3 \times 10^{16}$ GeV, and the
complexity we see at low energies arises from the running of the
renormalization group equations (RGEs) down to the electroweak
scale. This approach, however, has not appeared to be too
promising. Thus a general analysis of the $u$ and $d$ Yukawa
matrices with five zeros at the GUT scale given in \cite{rrr} is
shown in \mbox{Table 1}. Here $\lambda = 0.2$ is the Wolfenstein
parameter, and the choice of Table 1, when evaluated  at the
electroweak scale does indeed agree  approximately with the quark
masses and CKM matrix. However, to generate the experimental
hierarchy one has to have entries at the GUT scale of size
$\lambda^6 \simeq 10^{-5}$, showing that the problem at the GUT
scale is very much the same as at the electroweak scale.

\vspace{1cm} {Table 1. Approximate Yukawa textures at $M_G$ for
$Y_U$ and $Y_D$ where $\lambda$=0.2 \cite{rrr}.
 }\hrule{}\vspace{0.5cm}\begin{center}
$ Y_U=\left(\begin{matrix} 0 & \sqrt{2}\lambda^{6}  & 0 \cr
\sqrt{2}\lambda^{6} & \sqrt{3}\lambda^{4} & \lambda^{2} \cr 0  &
\lambda^{2}& 1 \end{matrix}\right); \,\,\,Y_D=\left(\begin{matrix}
0 &  2\lambda^{4}  & 0 \cr 2\lambda^{4} & 2\lambda^{3} & 0 \cr 0
& 0& 1 \end{matrix}\right).$
\end{center}
\hrule{}

\vspace{1cm}

String theory represents at present the only model that has been
proposed which in principle can calculate the Yukawa matrices from
first principles. Unfortunately, mathematical tools to explicitly
do this have not yet been developed. In spite of this, the general
formulation of the Yukawa problem in string theory opens new
windows for seeing how the quark and lepton hierarchies might
naturally have arisen, approaches not available in standard SUGRA
GUT theory. In particular, the Horava-Witten heterotic M-Theory
\cite{hw1,hw2}, which offers a natural explanation of why grand
unification can occur at $M_G$ rather than the Planck scale
$M_P$, has had significant development
\cite{w1,h1,bd,fmw1,fmw2,rd,low1,low2,And1,gc,losw,lpt,low3,dlow1,dlow2,low4,o,lo,dopw,blo,CK1,And2,CK2}
giving rise to three generation models with the SM low energy
gauge group $SU(3)\times SU(2)\times U(1)$. In this model,
physical space is one of two 10 dimensional (10D) orbifold planes
separated by a finite distance in the 11th dimension, the theory
obeying $S^1/Z_2$ symmetry in the 11th dimension. Six of the 10
dimensions are compactified to a Calabi-Yau (C-Y) threefold, the
remaining four being Minkowski space. An array of six
dimensional 5-branes perpendicular to the 11th dimension can exist
between the two orbifold planes.  While it is not possible to make
first principle calculations, one can examine whether the general
structure of such a  theory can replicate the SM at low energy. In
this connection, it was seen in \cite{ab} that the general
structure of the quark mass matrices can arise without undue fine
tuning if the 5-branes lie close to the distant orbifold plane,
and the instanton number of the physical orbifold plane, $\beta^{(0)}$ vanished.
It was explicitly shown in \cite{ab} that a three generation model
with $\beta^{(0)}=0$ and SM gauge group indeed can exist for a torus fibered
Calabi-Yau (with two sections) with del Pezzo base $dP_7$. The
quark and CKM matrix were calculated for a model of this type in
agreement with data, and it was shown also in approximate analytic
calculations how the mass hierarchies can arise without undue fine
tuning due to the general structure of the Kahler potential.

In this paper we extend the analysis of \cite{ab} in two
directions. We first include the charged lepton mass matrix and
obtain the mass hierarchies experimentally seen. We then consider
neutrino masses. The conventional way for accounting for the very
small mass of neutrinos is the seesaw mechanism \cite{seesaw}
which gives rise to Majorana neutrino masses. We consider here,
however, a new way of achieving small neutrino masses based on the
structure of the Kahler potential. This mechanism is different
from the seesaw mechanism, and gives rise to Dirac masses for the
neutrinos. Neutrino masses and the MNS matrix \cite{MNS} are
calculated consistent with the large mixing angle (LMA) analysis
of the solar, atmospheric, reactor and long baseline neutrino data
(e.g. see \cite{GG} for a global analysis in the context of
\mbox{three-neutrino} oscillations). Because of the appearance of the MNS
matrix, lepton flavor violation processes, which are absent in the
SM, will occur in this model. SM contributions here are too small
to be observed experimentally. However, SUSY contributions are
much larger. We have studied the $\muegamma$ decay in this model
including all possible contributions. For large $\tan\beta$, the
branching ratio for $\muegamma$ is close to the current
experimental bound \cite{PDG02a}, and would be accessible to
future experiments \cite{Mori99,MECO}. A summary of some of the
above results was given in \cite{ADH02}.

Our paper is organized as follows. In Sec.~2 we give a brief
review of M-Theory, and the basic results obtained in \cite{ab}
for torus fibered Calabi-Yau manifolds. In Sec.~3 we review and
update the results of \cite{ab} for the quark masses and extend
this analysis to the lepton sector. In Sec.~4 we introduce the new
mechanism to obtain small neutrino masses and calculate the masses
and mixing angles for this model. In Sec.~5 we present our
calculation for the $\muegamma$ decay. Finally, conclusions are
given in Sec.~6.

\section{Horava-Witten Kahler Potential}

The Horava-Witten M-Theory is concerned with 11 dimensional
supergravity on an orbifold $M_{10} \times S^1/Z_2$, where $Z_2$
is reflection of the 11th coordinate. One can think of this space
as an 11 dimensional space $M_{11}$ bounded by two 10 dimensional
orbifold planes $M_{10}$ at $x^{11} = 0$ and $\pi\rho$. In the
simplest case, $M_{10}$ is the product space $M_4 \times X$ where
$M_4$ is Minkowski space and $X$ is a (compact) C-Y threefold, the
physical world living on one of the orbifold planes (e.g. $x^{11}
= 0$), the other orbifold plane being a ``hidden'' sector. In
addition, there may be six dimensional 5-branes lying along
$x^{11}$ at bulk points $x_n$ with $0 < x_n < \pi\rho$, parallel
to the orbifold planes, each with four dimensions spanning $M_4$,
the additional two dimensions wrapped around a holomorphic curve
in the Calabi-Yau space. The construction of a consistent theory
involves a remarkable set of interlocking constraints due to
anomaly cancellation, gauge invariance, and local supersymmetry
leading naturally to a theory which possesses a number of
properties appropriate for phenomenology. Thus there must be $E_8$
gauge interactions with chiral multiplets on each $M_{10}$
orbifold plane (SO(32) being excluded) which can easily be broken on the physical plane 
to the SM group by Wilson lines. The 10D gauge coupling constant,
$\lambda$, is uniquely determined in terms of the 11D Planck mass,
$\kappa^{-2/9}$, leading to the result that the fundamental scale
of nature, the 11D Planck mass, is ${\cal O}(M_G)$, and explaining
why grand unification occurs at $M_G$ rather than the 4D Planck
mass (which is a derived quantity). Finally, a consistent theory
exists only as a quantum theory (the classical theory being
inconsistent), something one would hope might be true for any
fundamental theory. Much progress has been made in showing what
the low energy structure of such a theory might be, and models
with three generations of quarks and leptons obeying the SM gauge
group have been constructed. While the details of the construction
of the theory given in \cite{hw1,hw2} is rather intricate, it is
possible to see how the different elements interact to produce a
physically interesting model and so we first briefly summarize
this construction. We then give the relevant formulae needed to
examine the low energy structure. Details of the latter can be
found in \cite{low4}, and for the specific model considered here
in \cite{ab}.

The field content of 11D supergravity is the metric $g_{IJ}$, the
gravitino $\psi_{IJ}$, the three form $C_{IJK}$ and its field
strength $G_{IJKL}$. (In lowest order $G_{IJKL} = d_I C_{JKL}$.)
The Bose part of the Lagrangian is :
\begin{eqnarray}\label{eq201}
L_S & = & \frac{1}{\kappa^2}\int_{M^{11}} d^{11}x \sqrt{g}
          \left(-\frac{1}{2}R - \frac{1}{48}G_{IJKL}G^{IJKL} \right. \nonumber\\
    &   & \left. - \frac{\sqrt{2}}{3456}\epsilon^{I_1 I_2
          \ldots I_{11}}C_{I_1 I_2 I_3} G_{I_4\ldots I_7}G_{I_8 \ldots I_{11}}\right).
\end{eqnarray}
where the field strengths obey the field equations $D^IG_{IJKL} =
0$, and the Bianchi identity $dG_{IJKLM} = 0$. Here $\kappa$ is
the 11D gravitational constant. The Horava-Witten theory comes
about as follows. While in a smooth manifold 11D supergravity has
no anomalies, on an orbifold anomalies arise at the fixed points
$x=0$ and $x=\pi \rho$. To cancel these, it is necessary to put
Yang Mills multiplets on each $M_{10}$ orbifold plane, and the
cancellation occurs only if the gauge group on each manifold is
(the phenomenologically desirable) $E_8$. To lowest order, the Yang
Mills Lagrangian on each $M_{10}$ reads then:
\begin{equation}\label{eq202}
L_{YM}= -\frac{1}{\lambda^2} \int_{M^{10}} d^{10}x \sqrt{g}
\,\,\mathrm{tr} \left( \frac{1}{4} F_{AB} F^{AB} + \frac{1}{2}
\bar{\chi} \Gamma^{A} D_{A} \chi \right).
\end{equation}
where $A,B = 1,2 \ldots 10$, and $\chi$ is the associated gaugino.
However, Eq. (\ref{eq202}) is not locally supersymmetric, and one
must proceed in the usual fashion to add additional interactions
and modifications of the transformation laws to achieve local
supersymmetry. As usual, this involves coupling the gravitino
to the Yang Mills supercurrent. However, unlike the case where the
Yang Mills and supergravity multiplets live in the same space, the
gravitino here lives in the 11D bulk, while the Yang Mills
multiplet is constrained to live in 10D. For this situation, a
locally supersymmetric Yang Mills theory cannot be achieved simply
by adding interactions on the orbifold plane. It turns out that a
supersymmetric theory can be achieved only by modifying the
Bianchi identities to read
\begin{equation}\label{eq203}
d G_{11 \, ABCD} = 8\pi^2\sqrt{2}\frac{\kappa^2}{\lambda^2}
\Sigma_0^{N+1} J^{(n)} \delta (x^{11}-x_n).
\end{equation}
where $x_0 = 0$, $x_{N+1} = \pi \rho$ and $x_n$, $n= 1\ldots N$
are the positions of the five branes,
\begin{equation}\label{eq204}
J^{(0,N+1)} = -\frac{1}{16 \pi^2}\left( \mathrm{tr}\,F\wedge F -
\frac{1}{2}\,\mathrm{tr}\,R\wedge R \right)_{x^{11}=0, \pi\rho}.
\end{equation}
and $J^{(n)}$, $n= 1\ldots N$ are sources from the 5-branes. With
Eq.(\ref{eq203}), the total supergravity + (modified) Yang Mills
Lagrangian can be made locally supersymmetric. However, having
gained supersymmetry, one has lost Yang Mills gauge invariance
(!). For while Eq.(3) implies that $G_{ABCD}$ is gauge invariant,
the corresponding potential $C_{11\,AB}$ now is not, i.e. under a
Yang Mills gauge transformation one has
\begin{equation}\label{eq205}
\delta C_{11\, AB} = - \frac{\kappa^2}{6\sqrt{2}\lambda^2}\left[
\mathrm{tr} \left( \epsilon F_{AB} \delta(x^{11}) \right) +
\mathrm{tr}\left( \epsilon F_{AB} \delta(x^{11} - \pi\rho) \right)
\right].
\end{equation}
which implies the $C\wedge G\wedge G$ term of Eq.(\ref{eq201}) is
not gauge invariant. Thus the classical theory is not gauge
invariant, and a consistent classical theory does not exist.
However, in the quantum theory, there is in addition the 10D
Majorana-Weyl anomaly, and due to unique features of the $E_8$
group (!) can cancel the loss of gauge invariance of the
``Green-Schwarz" $C\wedge G\wedge G$ term provided
\begin{equation}\label{eq206}
\lambda^2 = 2 \pi (4\pi\kappa)^{2/3}.
\end{equation}
Thus only a consistent quantum theory can be built, and this
quantum theory determines the 10D gauge coupling constant in terms
of the  11D gravitational constant.

Eq.(\ref{eq206}) leads immediately to interesting phenomenological
consequences. For compactifying $M_{11}$ on a \mbox{Calabi-Yau} manifold,
one has to lowest order for the 4D gauge coupling constant and
Newton constant \cite{w1}
\begin{equation}\label{eq207}
\alpha_G=\frac{(4 \pi \kappa^2)^{2/3}}{2 \mathcal{V}};\,\,\,\,
G_N=\frac{\kappa^2}{16 \pi^2 \mathcal{V}\rho}
\end{equation}
where $\mathcal{V}$ is the Calabi-Yau volume. Setting
$\mathcal{V}^{1/6} = 1/M_G$ (so that grand unification occurs at
the compactification scale as required by the LEP data) and using
$\alpha_G = 1/24$, one finds that the fundamental 11D Planck mass
is $\kappa^{- 2/9} \cong 2M_G$ and $\pi\rho^{-1} \cong 4.7 \times
10^{15}$ GeV. Alternately one may say that the 11D Planck mass is
the fundamental scale and it sets the GUT scale, while the
largeness of the 4D Planck mass is due mostly to accidental $4\pi$
factors arising in the analysis.

We now summarize the basic formulae of \cite{low4} and \cite{ab}
needed to build a phenomenologically acceptable theory. The
sources $J^{(n)}$ of Eq. (\ref{eq203}) play an important role in
building a model. Thus if integrated over a set of independent 4
cycles $C_{4i}$, they define integer charges:
\begin{equation}\label{208}
\beta_i^{(n)} = \int_{C_{4i}} J^{(n)}
\end{equation}
and Eq.(\ref{eq203}) then implies $\Sigma\beta_i^{(n)} = 0$. Here
$\beta_i^{(0)}$ and $\beta_i^{(N+1)}$ are the instanton charges on the orbifold  
planes and $\beta_i^{(n)}$ ($n=1\ldots N$) are the magnetic charges of the 
5-branes. The 
existence of non-zero instanton Yang Mills fields with gauge group
G on the orbifold plane implies that $E_8$ breaks into $G\times H$
where H is the remaining symmetry at the GUT scale of the physical
theory. We chose here $G = SU(5)$ so that  $H =
SU(5)$\footnote{Recently, an alternate choice, $G = SU(4)$ and $H
= SO(10)$ has been considered by A.~E.~Faraggi and R.~S.~Garavuso
\cite{Far1, Far}.}.

Chiral matter arises from the components of the Yang Mills
multiplet in the \mbox{Calabi-Yau} part of the $M_{10}$ orbifold
\cite{low4}.  Thus labeling the C-Y indices by holomorphic
(\mbox{anti-holomorphic}) coordinates $a (\bar{a}) = 1,2,3$, then one can
expand e.g. $F_{\mu \bar{b}}$ in terms of a basis set of functions
$u_I^x$ in the C-Y space (I is a family index and x a
representation index), the coefficients in the Minkowski space
being the scalar components of the chiral multiplets $C(R)^{Ip}$ 
(where R is the representation):
\begin{equation}\label{eq209}
F_{\mu\bar{b}} = \sqrt{2\pi\alpha_{G}}\, \sum_R
u^x_{I\bar{b}}(R)\, T_{xp}(R) (D_\mu C(R))^{Ip}.
\end{equation}
In terms of these quantities, one then defines the metric
\begin{equation} \label{eq210}
G_{IJ}(a^i;R)=\frac{1}{{vV}}\int_X\sqrt{g}g^{a\bar
b}u_{Iax}(R)u^x_{J\bar b}(R)
\end{equation}
and the Yukawa couplings \cite{low4}
\begin{equation}\label{eq211}
\lambda_{IJK}(R_1,R_2,R_3)=\int_X\Omega\wedge u^x_{I}(R_1)\wedge
u^y_{J}(R_2)\wedge u^z_{K}(R_3)f_{x,y,z}^{(R_1,R_2,R_3)}
\end{equation}
where $\Omega$ is the covariantly constant (3,0) form, f projects
out the gauge singlet parts, and ${\mathcal{V}}\equiv vV$ is the
volume of the Calabi-Yau space while $v$ is the coordinate volume:
\begin{equation}\label{eq212}
V=\frac{1}{v}\int_Xd^6x\sqrt g;\,\,\,v=\int_Xd^6x
\end{equation}
In addition one defines the $S$, $T^i$ and 5-brane moduli by
\begin{equation}\label{eq213}
Re(S)=V;\,\, Re T^i=V^{-1/3}Ra^i;\,\, Re Z_n=z_n
\end{equation}
where the modulus $R$ is the orbifold radius divided by $\rho$ and
$z_n=x_n/{\pi\rho}$.  $V$ can be expressed in terms of the $a^i$
moduli by $V(a)=\frac{1}{6}d_{ijk}a^ia^ja^k$ where $d_{ijk}$ are
the Calabi-Yau intersection numbers :
\begin{equation}\label{eq214}
d_{ijk}=\int_X\omega_i\wedge\omega_j\wedge\omega_k
\end{equation}
Following the techniques of \cite{w1}, the field equations and
Bianchi identities in Eq.(\ref{eq203}) were solved in the presence
of 5-branes to leading order $O(\kappa^{2/3})$ \cite{low4} leading
to an effective four dimensional Lagrangian at compactification
scale $M_G$. We now state the results that were obtained. The
gauge kinetic functions on the orbifold planes are given by
\begin{eqnarray}\label{eq215}
f^{(1)}&=&S+\epsilon T^i \left
(\beta^{(0)}_i+\sum^{N}_{n=1}(1-Z_n)^2\beta^{(n)}_i \right )
\nonumber\\ f^{(2)}&=&S+\epsilon T^i\left
(\beta^{(N+1)}_i+\sum^{N}_{n=1} Z_n^2\beta^{(n)}_i \right)
\end{eqnarray}
where
\begin{equation}\label{eq216}
\epsilon=\left(\frac{\kappa}{{4\pi}}\right)^{2/3}\frac{2\pi^2\rho}{\mathcal{V}^{2/3}}
\end{equation}
The matter Kahler potential, $K=Z_{IJ}\bar {C^I}C^J$, on the
physical orbifold plane at $x^{11}=0$ has the Kahler metric
\begin{equation}\label{eq217}
Z_{IJ}=e^{-K_T/3}\left [G_{IJ}-\frac{\epsilon}{2V}\tilde{\Gamma}^i_{IJ}
\sum^{N+1}_{n=0}(1-z_n)^2\beta^{(n)}_i \right ]
\end{equation}
where
\begin{equation}\label{eq218}
K_T=-\ln
[\frac{1}{6}d_{ijk}(T^i+\bar{T^i})(T^j+\bar{T^j})(T^k+\bar{T^k})]
\end{equation}
\begin{equation}\label{eq219}
\tilde{\Gamma}^i_{IJ}=\Gamma^i_{IJ}-(T^i+\bar{T^i})G_{IJ}-\frac{2}{3}
(T^i+\bar{T^i})(T^k+\bar{T^k})K_{Tkj}\Gamma^j_{IJ}
\end{equation}
and
\begin{equation}\label{eq220}
K_{Tij}=\frac{\partial^2K_T}{\partial
T_i\partial\bar{T^{j}}};\,\,\Gamma^i_{IJ}= K^{ij}_T\frac{\partial
G_{IJ}}{\partial T^j}
\end{equation}
The Yukawa matrices are
\begin{equation}\label{eq221}
Y_{IJK}=2\sqrt{2\pi\alpha_G}\lambda_{IJK}\simeq 1.02\lambda_{IJK}
\end{equation}
for $\alpha_G=1/24$. The Kahler metric on the distant orbifold
plane at $x^{11}=\pi\rho$ is given by Eq.(\ref{eq217}) with
$z_n\rightarrow(1-z_n)$.

\section{Yukawa Textures}
The Yukawa couplings are given in Eqs.(\ref{eq211}) and
(\ref{eq221}) as integrals over the C-Y space. A priori there is
no reason to suggest that a hierarchy such as Table 1 should arise
and one expects that the non-zero entries to be $O(1)$. Similarly,
one expects a priori that the non-zero elements of $G_{IJ}$ in
Eq.(\ref{eq210}) be of $O(1)$. However, a mild hierarchy can
develop in the Kahler metric of Eq. (\ref{eq217}) if the 5-branes
all lie close to the distant orbifold plane, i. e. $d_n = 1 - z_n
\cong 0.1$, and provided also $\beta^{(0)} = 0$. Then the second
term will be small compared to the first ($\epsilon \cong 0.9$),
and the model of \cite{ab} assumed that $G_{IJ}$ contributes only
to the first two generations of the $u$ quark and $d_L$ (which
appear together in the $SU(5)$ 10 representation) but to all
generations of $d_R$, while the second term contributes to all
generations but is then dominant for the third generation of
$u_L$, $u_R$, $d_L$. (That a C-Y manifold exits with $\beta^{(0)}
= 0$ with three generations and a SM gauge group is non-trivial
and was explicitly shown to be possible in \cite{ab}.)  When the
Kahler metric was diagonalized to a unit matrix, it was seen that
this idea was sufficient to generate a satisfactory explanation of
the more extreme Yukawa hierarchies at the electroweak scale, and
we extend this idea here to the lepton sector. Thus the Kahler
metric has the general form
\begin{equation}\label{eq301}
    Z^F=f_T\left(\begin{matrix} 1  & O(1)  & O(d^2) \cr O(1) & O(1) & O(d^2)
\cr O(d^2)  & O(d^2) & O(d^2) \end{matrix}\right)
\end{equation}
where F stands for the different matter fields: $q = u_L, u_R,
d_L$, $l = (\nu_L, e_L)$ and $e = e_R$ and $f_T$ is given from
Eq.(\ref{eq217}) to be $e^{-K_T/3}$. We assume that $G_{IJ}$ has
non-zero elements of $O(1)$ for all generations of $d_R$.
(For convenience, we've re-scaled the $Z^F_{11}$ entry in Eq. (22)
to 1.) The hierarchy then arises when one transforms the $Z_{IJ}$
to the unit matrix by a unitary matrix $U$ and a diagonal scaling
matrix $S$ to obtain the canonical matter fields
${C^I_F}^{\prime}$:
\begin{equation} \label{eq302}
C^I_{F}=\frac{1}{\sqrt{f_T}}(U^{(F)} S^{(F)})_{IJ}
{C_{F}^{J}}^{\prime}
\end{equation}
where
\begin{equation} \label{eq303}
{\rm diag} S^{(F)}= (\lambda_{F1}^{-1/2},\,
\lambda_{F2}^{-1/2},\,\lambda_{F3}^{-1/2}).
\end{equation}
and $\lambda_{Fi}$, $i = 1,2,3$ are the eigenvalues of
$Z^F_{IJ}/f_T$. A similar transformation is made on the Higgs
fields contribution to the Kahler potential
\begin{equation}\label{eq304}
 f_T G_{H_{1,2}}{\bar {H}}_{1,2}H_{1,2}
\end{equation}
with rescaling of $H_{1,2}$:
\begin{equation}\label{eq305}
H_{1,2}=\frac{1}{\sqrt{f_T G_{H_{1,2}}}}H^{\prime}_{1,2}
\end{equation}

Before making the transformation of Eq. (\ref{eq302}), The Yukawa
contribution to the superpotential is \cite{low4}
\begin{equation}\label{eq306}
W_Y=e^{\frac{1}{2}K_m}\frac{1}{3}Y_{IJK}C^IC^JC^K
\end{equation}
where $K_m=\ln(S+\bar{S})+K_T$ is the moduli contribution to the
Kahler potential. From Eqs. (\ref{eq213}) and (\ref{eq218}), one
has
\begin{equation}\label{eq307}
K_m=-ln(2 V)-ln({8}R^3).
\end{equation}
Written in terms of SM fields $W_Y$ then is
\begin{equation}\label{eq308}
W_Y=\frac{1}{4R^{3/2}V^{1/2}}(Y^{(u)} q_L H_2 u_R + Y^{(d)} q_L
H_1 d_R + Y^{(e)} l_L H_1 e_R).
\end{equation}
and after the transformation to the canonical matter fields one
has
\begin{equation}\label{eq309}
W_Y=u_L^{\prime} \lambda^{(u)} u_R^{\prime} H_2^{\prime} +
d_L^{\prime} \lambda^{(d)} d_R^{\prime} H_1^{\prime} +
e_L^{\prime} \lambda^{(e)} e_R^{\prime} H_1^{\prime}.
\end{equation}
where $\lambda^{(u,d,e)}$ are give by \footnote{We correct an
error in \cite{ab}, the omission of the
$V^{-\frac{1}{3}}$ factor in Eq.(\ref{eq213}) (see e.g. \cite{Lima}),  which leads to a
factor $1/V^{-\frac{1}{2}}$ in Eq.(\ref{eq310}-\ref{eq312}) rather
then $1/V$.}
\begin{equation}\label{eq310}
    \lambda^{(u)}_{IJ} = \frac{1}{8 \sqrt{2}}\frac{1}{R^3V^{1/2}}\frac{1}{\sqrt{G_{H_2}}}
                      (S^{(q)}{\tilde U^{(q)}}Y^{(u)}U^{(u)}S^{(u)})_{IJ}
\end{equation}

\begin{equation}\label{eq311}
    \lambda^{(d)}_{IJ} = \frac{1}{8 \sqrt{2}}\frac{1}{R^3V^{1/2}}\frac{1}{\sqrt{G_{H_1}}}
                      (S^{(q)}{\tilde U^{(q)}}Y^{(d)}U^{(d)}S^{(d)})_{IJ}
\end{equation}

\begin{equation}\label{eq312}
    \lambda^{(e)}_{IJ} = \frac{1}{8 \sqrt{2}}\frac{1}{R^3V^{1/2}}\frac{1}{\sqrt{G_{H_1}}}
                       (S^{(l)}{\tilde U^{(l)}}Y^{(e)}U^{(e)}S^{(e)})_{IJ}
\end{equation}
We use here the notation ``$\sim$" for transpose. In Eq.
(\ref{eq309}), $\lambda^{(u,d,e)}$ play the role of the Yukawa
matrices at the GUT scale in the phenomenological analyses such as
in \cite{rrr}. However, in general they are not symmetric matrices
and so M-Theory textures are uniquely different from what has
previously been considered in phenomenological analyses. In brief,
it is the smallness of the third generation eigenvalues of the
Kahler matrices appearing in the denominators of
Eq.(\ref{eq310}-\ref{eq312}) (from the factor S of
Eq.(\ref{eq303})) that give rise to the large third generation
masses.
\begin{table}
\begin{flushleft}{Table 2. Kahler matrices $Z^{(u,d,l,e)}_{IJ}$ and
Yukawa matrices  $Y^{(u,d,e)}$ for $\tan\beta$=40.}\end{flushleft}
\vspace{0.2cm} \hrule{}
\begin{eqnarray}\nonumber
Z^u=f_T\left(\begin{matrix} 1  & 0.3452  & 0 \cr 0.3452 & 0.1311 &
0.006365 \cr 0  & 0.006365 & 0.00344 \end{matrix}\right); \,\,\,
Z^d=f_T\left(\begin{matrix} 1 & 0.496  & 0 \cr 0.496 & 0.564 &
0.435 \cr 0  & 0.435 & 0.729  \end{matrix}\right);
\end{eqnarray}
\begin{eqnarray}\nonumber
Z^l=f_T\left(\begin{matrix} 1  & -0.547  & 0 \cr -0.547 & 0.432 &
0.025 \cr 0  & 0.025 & 0.09 \end{matrix}\right); \,\,\,
Z^e=f_T\left(\begin{matrix} 1 & 0.624  & 0 \cr 0.624 & 0.397 &
0.00574 \cr 0  & 0.00574 & 0.004407  \end{matrix}\right).
\end{eqnarray}
\vspace{0.2cm}
\begin{eqnarray}\nonumber
{\rm diag}Y^{(u)}&=&{(0.0114,\, 0.0597,\, 0.104\mspace{1mu}\exp[0.65 \pi i])};
\\\nonumber
{\rm diag}Y^{(d)}&=&{(2.052,\, 0.2565,\, 1.8297)};
\\\nonumber
{\rm diag}Y^{(e)}&=&{(0.307,\, 3.789,\, 1.821)}.
\end{eqnarray}
\hrule{}\end{table}
\begin{table}
\begin{flushleft}{ Table 3. Quarks and leptons masses and CKM matrix
elements obtained from the model of Table 2. Masses are in GeV.
Experimental values for lepton and quark masses are from \cite{PDG02a}
and CKM entries from \cite{lp03} unless otherwise noted.}\end{flushleft}
\vspace{0.2cm} \hrule{}
\begin{center} \begin{tabular}{|c|c|c|}
 \hline Quantity&Theoretical Value&Experimental Value\\\hline
$m_t$(pole)& 175.2 & $174.3 \pm5.1$ \\
$m_c$($m_c$)& 1.27 & 1.0-1.4\cr
$m_u$(1 GeV)& 0.00326 & 0.002-0.006\\
$m_b$($m_b$)& 4.21 & 4.0-4.5\cr
$m_s$(1 GeV)& 0.086 & 0.108-0.209\\
$m_d$(1 GeV)& 0.00627 & 0.006-0.012\\
$m_\tau$& 1.78 & 1.777\\
$m_\mu$& 0.1054 & 0.1056\\
$m_e$& 0.000512 & 0.000511\\
$|V_{us}|$& 0.221 & $0.2210\pm 0.0023$ \\
$|V_{cb}|$& 0.042 & 0.0415$\pm$0.0011 \\
$|V_{ub}|$& $4.96 \times 10^{-3}$ & $3.80^{+0.24}_{-0.13} \pm 0.45 \times 10^{-3}$ \\
$|V_{td}|$& $6 \times 10^{-3}$ & $9.2 \pm 1.4 \pm 0.5 \times 10^{-3}$ \\
$\sin 2\beta$ & 0.803 &  $0.731\pm 0.056$ \cite{hfag}\\
\hline\end{tabular}\end{center} \hrule{}\vspace{1cm}
\end{table}

In \cite{ab} we saw for the case of $\tan\beta = 3$ how the above
Yukawa matrices gave rise to the experimental quark masses and CKM
matrix elements at the electroweak scale, and we showed there
analytically how the hierarchies arose naturally without undue
fine tuning. We now update this analysis for the case of
$\tan\beta = 40$, and extend the discussion to include the lepton
sector. Table 2 shows a choice of Kahler metric and Yukawa matrices
that satisfy all the current experimental data. The $Z^F_{23}$, $Z^F_{32}$ and
$Z^F_{33}$ entries for $F=u,l,e$ are $O(d^2)$ (for $d=0.1$) as
required by Eq.(\ref{eq301}). For simplicity we have assumed that
the $q$ and $u$ quarks have identical Kahler matrices and have the
maximum number of zero entries, and that the Yukawa matrices are
diagonal. One phase is assumed in the Yukawa matrices to account
for CP violation. To compare with low energy data, we use one loop
Yukawa RGEs and two loop gauge RGEs to evaluate the Yukawa
couplings at the electroweak scale, which we take to be $m_t$.
Below $m_t$ we assume that the Standard Model holds and include in
our calculations the QCD corrections (which are quite
significant). The QCD correction factors used were $\eta_c = 2$,
$\eta_u = 2.5 = \eta_d$, $\eta_b=1.6$ and $\eta_s=2.5$. Diagonalization of the low energy Yukawa
matrices then allows one to generate the low energy quark and
lepton masses and the CKM matrix elements. The results are shown
in Table 3, and are in good agreement with experiment. Of course
in a fundamental analysis, the precise entries in Table 2 arise
from integrals over the Calabi-Yau space, an analysis that cannot
at this stage be performed. However, our discussion has shown that
the general structure of the Kahler metric and Yukawa couplings
arising in our Horava-Witten model can lead to low energy quark
and lepton spectra consistent with all current experiments without
the fine tuning used in phenomenological analyses.

Without knowledge of the value of the factors
$R^3V^{1/2}\sqrt{G_{H_{1,2}}}$ in the denominators of
Eq.(\ref{eq310}-\ref{eq312}), Kahler textures can only determine
the mass ratios. As in \cite{ab}, we use the top Yukawa at the GUT
scale to determine the value of this common factor. If we write
$V=r^6$, where r is the mean radius of the Calabi-Yau manifold
divided by the co-ordinate radius, then for $G_{H_{1,2}}=1$, one
finds that
\begin{equation}\label{eq312a}
    R \times r = 6.82.
\end{equation}
In the next section, we will show that $R$ and $r$ can be
determined separately if massive neutrinos enter our model via the
mechanism proposed there.

\section{Neutrino masses and Oscillations}
In the last section we presented a way to generate the Yukawa
textures in the quark and lepton sectors whose structures are the
same as the SM. The consequence of the masslessness of neutrinos
in the SM is that the mass eigenstates of leptons are identical to
their gauge or flavor eigenstates and, unlike the quark sector
which has a CKM mixing matrix, the lepton sector does not.
Therefore, there is no oscillations between neutrinos in the SM.
However, the neutrino experiments of Super-Kamiokande \cite{SK,solar},
SNO \cite{SNO} and KamLAND \cite{Kamland} have shown the existence
of neutrino oscillations which indicates that neutrinos are
actually massive particles. In this section we will show that
massive neutrinos can be included in our model and their masses
and mixings can be fitted into the large mixing angle (LMA)
solution. (For a recent review of neutrino oscillations see
\cite{Bilenky03a}.)

The simplest way to include massive neutrinos to our model is to
associate a right-handed neutrino to every left-handed neutrino
and insert by hand a term proportional to
\begin{equation}\label{eq401}
Y^{(\nu)} l_L H_2 \nu_R
\end{equation}
into superpotential (\ref{eq308}). However, the Yukawa couplings
in the neutrino sector have to be extremely small and thus this
solution is theoretically less interesting unless there is a
mechanism behind it. The most widely used way to overcome this
problem is the seesaw mechanism \cite{seesaw}. In seesaw models,
besides the usual Dirac mass terms (which are approximately the
same size as other fermion masses), one introduces additional very
large Majorana masses which enter in the off-diagonal entries of
the neutrino mass matrix. As a consequence, some eigenvalues are
suppressed to the desired values when the diagonalization of
neutrino mass matrix takes place. The physical neutrinos in seesaw
models are then of Majorana type while other leptons and quarks
are Dirac fermions. Here we propose a new way to generate neutrino
masses. In our model, neutrinos are of Dirac type and thus the
similarity between leptons and quarks is preserved and no
neutrinoless double beta decay exists. We will see that our new
mechanism provides a reasonable physical explanation to the origin
of term (\ref{eq401}).

The Kahler potential in principle can have gravitationally coupled
trilinear terms which are usually ignored as they generally are of
negligible size. However, we assume here that our Kahler potential
at the GUT scale contains the holomorphic cubic term
$K^{(3)}=K_\nu + {K_\nu}^{\dag}$ where
\begin{equation}\label{eq402}
K_{\nu}=\kappa_{11} Y^{(\nu)} l_L H_2 \nu_R
\end{equation}
where $1/\kappa_{11}$ is the 11 dimensional Planck mass (i.e.
$1/\kappa_{11}\simeq M_G$) and $Y^{\nu}$ is a Yukawa matrix. We note
that Eq.(\ref{eq402}) is the only gauge invariant holomorphic
cubic lepton term involving $\nu_R$ and that $\kappa_{11}$ is the
natural scale for Horava-Witten theory. We assume here that the Yukawa 
contribution to the superpotential is still given by (\ref{eq309}),
and that no additional neutrino masses arise there. One can transfer 
$K_{\nu}$ from the Kahler potential to the superpotential by a Kahler
transformation ($1/\kappa_4$ is the 4D Planck mass):
\begin{eqnarray} \label{eq403}
K & \rightarrow  & K - K^{(3)},  \nonumber \\
W & \rightarrow  & e^{\kappa_4^{\;2} K_{\nu}} W = W + 
\kappa_4^{\;2} K_{\nu} W + \cdots
\end{eqnarray}
Now when supersymmetry breaks, the superpotential W will grow a
VEV of size:
\begin{equation}\label{eq404}
\left\langle W \right\rangle \cong \frac{1}{\kappa_4^{\;2}} M_S
\end{equation}
where $M_S$ is of electroweak size. Consequently, after
supersymmetry breaking, an additional term appears in
superpotential (\ref{eq309}):
\begin{equation}\label{eq405}
\frac{M_S}{M_G} Y^{(\nu)} l_L H_2 \nu_R.
\end{equation}

In the above we have assumed that no additional neutrino masses
arise in the superpotential. One might imagine that this could come
about if there exists a global symmetry or a non-gauge discrete symmetry
that $\nu_R$ obeys. (Examples might be if $\nu_R$ was a member or a global
$SU(2)_R$ doublet or were charged under a global $U(1)$ symmetry; or if
$\nu_R$ appeared in $W$ only as an even power one could assign it a discrete
quantum number $-1$.) Then in the cubic part of the superpotential, the global
or discrete symmetry (plus $SU(2)_L$ symmetry) would forbid Majorana or Dirac
masses formed from $\nu_{L,R}$. Since gravity is expected to break such global or
discrete symmetries in string theories, quartic terms scaled by $\kappa_{11}$ could
arise (which is in fact why the contribution of Eq.~(\ref{eq402}) would be allowed
in the Kahler potential in this scenario.) Such terms would have the general form 
$(H_2 l)^2$  and $H_1H_2 \nu_R^2$, giving rise to Majorana masses. Using the 
parameters of Table 2 and Eq.~(\ref{eq409}) below, we estimate that the first term 
gives rise to neutrino masses a factor of $\sim 100$ smaller than Eq.~(\ref{eq402}), and the second term is a factor of $1/\tan\beta$ smaller yet. 
Thus they would produce only small corrections in our model. We have not
investigated in detail if one could dispense with Eq.~(\ref{eq402}) completely 
and try to get the correct neutrino masses using only the above superpotential
terms, but because of the above result, we believe it to be unlikely. We note also, 
that the above superpotential terms could also occur in the standard 
see-saw model, but are neglected there. While global and discrete symmetries can
indeed arise from Calabi-Yau manifolds in Horava-Witted M-theory, whether or not the
necessary symmetries are present in physically interesting Calabi-Yau manifolds is
not known. Thus we have postulated their existence in this paper.\footnote{We thank the Referee for bring this point to our attention.}

We can now proceed as in Sec.~2. First diagonalize and rescale the
Kahler matrices $Z_{IJ}$ of $\nu_R$ and other fields to the unit
matrix. Then make the necessary transformations in the
superpotential to the canonical normalized fields. The term giving
rise to neutrino masses can then be written as
\begin{equation}\label{eq407}
\nu_L^{\prime} \lambda^{(\nu)} \nu_R^{\prime} H_2^{\prime}
\end{equation}
where
\begin{equation}\label{eq408}
    \lambda^{(\nu)}_{IJ} = \frac{1}{\sqrt{2}}\frac{1}{R^{3/2}}\frac{1}{\sqrt{G_{H_2}}}\frac{M_S}{M_G}
                     (S^{(l)}{\tilde U^{(l)}}Y^{(\nu)}U^{(\nu)}S^{(\nu)})_{IJ}
\end{equation}
Note that the overall coefficient in (\ref{eq408}) is different
from the one in (\ref{eq310}-\ref{eq312}) because the neutrino
term originates from the Kahler potential, not the superpotential
(\ref{eq306}) which has the additional  coefficient
$e^{\frac{1}{2}K_m}$. It is thus possible to use the experimental
neutrino mass square differences to determine $R$. In the example
given below, we find that $R=2.13$ produces acceptable neutrino
masses (we assume $M_S=1$ TeV in our calculation), and from
Eq.(\ref{eq312a}), one finds that $r=3.20$. At the weak scale,
after the diagonalization of charged lepton and neutrino Yukawa
matrices, the Maki-Nakagawa-Sakata (MNS) lepton mixing matrix
arises. We follow the standard parameterization \cite{PDG02a} (the
phase similar to the one in the CKM matrix is ignored):

\begin{equation}\label{eq408a}
V_{MNS}=\left(\begin{matrix} c_{12}c_{13}  & s_{12}c_{13}  &
s_{13} \cr -s_{12}c_{23}-c_{12}s_{23}s_{13} &
c_{12}c_{23}-s_{12}s_{23}s_{13} & s_{23}c_{13} \cr
s_{12}s_{23}-c_{12}c_{23}s_{13}  &
-c_{12}s_{23}-s_{12}c_{23}s_{13} & c_{23}c_{13}
\end{matrix}\right).
\end{equation}
where $c_{ij}=\cos\theta_{ij}$, $s_{ij}=\sin\theta_{ij}$ and
$i,j=1,2,3$.

The following is an example at $\tan\beta=40$. We use the lepton
entries of Table 2, and the following neutrino Kahler and Yukawa
matrices at $M_G$:
\begin{eqnarray}\label{eq409}
Z^{\nu}=f_T\left(\begin{matrix} 1  & -0.465  & 0 \cr -0.465 &
0.3105 & 0.0254 \cr 0  & 0.0254 & 0.027 \end{matrix}\right);
\end{eqnarray}
\begin{eqnarray}\label{eq410}
{\rm diag}Y^{(\nu)}&=&{(4,\, 0.4,\, 4)}.
\end{eqnarray}
The neutrino mass square differences and mixing angles at the weak
scale are then calculated to be:
\begin{eqnarray}\label{eq411}
\Delta m_{21}^2&=& 5.5 \times 10^{(-5)} \;\; \text{eV}^2; \\
\Delta m_{32}^2&=& 2.7 \times 10^{(-3)} \;\; \text{eV}^2;
\end{eqnarray}
\begin{equation}\label{eq412}
\tan^2 \theta_{12} = 0.42; \;\; \tan^2 \theta_{23} = 0.93.\\
\end{equation}
with $|U_{e3}|=0.005$. Since our model is a complete model of
neutrino masses, we can calculate all the masses themselves and
not just the mass square differences. For the above example we
find \beq\label{eq412a} m_1 = 6.5 \times 10^{-4}
\;\text{eV};\;\;m_2 = 7.4 \times 10^{-3} \;\text{eV};\;\;m_3 = 5.2
\times 10^{-2} \;\text{eV} \eeq consistent with cosmological
constraints on neutrino masses \cite{wmap}.

The analysis of solar and KamLAND data in terms of two neutrino
oscillations gives for the LMA solution \cite{solar}:

\begin{equation}\label{eq413}
0.20\le \tan^2\theta_S \le 0.68\,;\;\;\;\;5.6\times 10^{-5}\le
\Delta m_S^2/\text{eV}^2\le 8.9\times 10^{-5}
\end{equation}
where $\Delta m_S^2$ is the solar neutrino
mass square difference and $\theta_S$ is the corresponding mixing
angle and the ranges in Eq.(\ref{eq413}) (and Eq.(\ref{eq414}) below) are $3\sigma$ around the central value. The analysis of
Super-Kamiokande and K2K data shows for the LMA solution \cite{atom}:

\begin{equation}\label{eq414}
0.85\le \sin^2 2\theta_A \le 1\,;\;\;\;\;1.4\times 10^{-3}\le
\Delta m_A^2/\text{eV}^2\le 3.8\times 10^{-3}
\end{equation}
where $\Delta m_A^2$ and $\theta_A$ are the
relevant mass square difference and mixing angle for the
atmospheric neutrino oscillation.

Since in our case $|U_{e3}|\cong0$, solar and atmospheric neutrino
oscillations decouple \cite{GG, BG}. Therefore the two neutrino
oscillation analysis can be applied to our case with the effective
mixing angles given by:
\begin{equation}\label{eq415}
\theta_S=\theta_{12}\, ,\;\;\theta_A=\theta_{23}.
\end{equation}
Eq.(\ref{eq411})-(\ref{eq412}), (\ref{eq413}) and (\ref{eq414})
show that our results agree with the current LMA solution quite
well.

\section{$\muegamma$ Decay}

Lepton flavor violation (LFV) processes in supersymmetric models
have been discussed in much detail in the literature (e.g.
\cite{CGR, PM, CK, BGOO, BDQ}). In our model the MNS matrix by
itself can give rise to LFV processes, but this contribution is
still very small, e.g. for $\muegamma$, the decay rate is of
${\cal O}{((m_\nu/m_W)^4)}$ \cite{Pet, Mar, LPSS, LS}. Therefore,
In this section, we mainly discuss the additional supersymmetric
contributions to the LFV process $\muegamma$.

The operator for $\muegamma$ is:
\begin{equation}\label{eq501}
{\cal L}_{\muegamma} = \frac{i e}{2 m_\mu} \, \overline{e}
\,\sigma^{\mu\nu} q_{\nu} \left( a_l P_L + a_r P_R \right) \mu
\cdot A_\mu + h.c.
\end{equation}
where $P_{L,R} \equiv (1 \mp \gamma_5)/2$ and $\sigma^{\mu\nu}
\equiv \textstyle{\frac{i}{2}} \,[ \gamma^{\mu}, \gamma^{\nu} ]$.
The decay width for $\muegamma$ can be written as:
\begin{equation}\label{eq502}
\Gamma(\muegamma) = \frac{m_\mu e^2}{64\pi} \left( |a_l|^2 +
|a_r|^2 \right)
\end{equation}
Then the branching ratio is given by:
\begin{eqnarray}\label{eq503}
\mbox{Br}(\muegamma) & \cong & \frac{\Gamma(\muegamma)}{\Gamma(\mu\rightarrow e \bar{\nu}_e \nu_\mu)} \nonumber \\
              & = & \frac{3 \pi^2 e^2}{G_F^2 m_\mu^4} \left( |a_l|^2 + |a_r|^2 \right)
\end{eqnarray}

\begin{figure}[t]
\begin{center}
\scalebox{1}[1]{\includegraphics{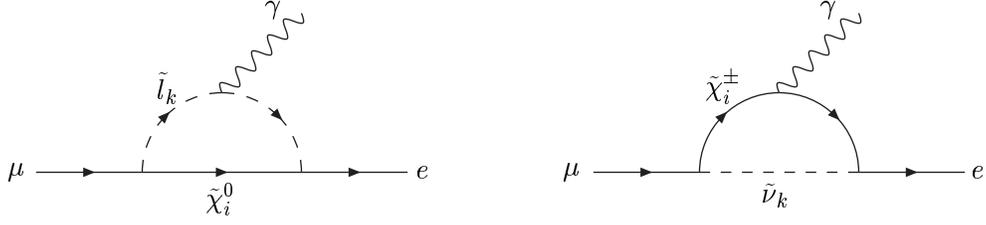}}
\end{center}
\caption{\label{fig1} Feynman diagrams for the neutralino and chargino contributions to $\mu\rightarrow e+ \gamma$.}
\end{figure}

The supersymmetric contributions include the neutralino and
chargino diagrams shown in FIG.\ref{fig1}.
The neutralino diagram gives: \beqa\label{eq504}
a^N_l & = & \sum_{i=1}^4 \sum_{k=1}^6 \frac{m^2_{\mu}}{8 \pi^2 m^2_{\tilde{l}_k}}
            \left( {N_{1L}^{ki}}^\ast{N_{2 L}^{ki}} \mbox{F}_1(x^N_{ik}) + \frac{m_{\tilde{\chi}^{0}_i}}
            {m_{\mu}}{N_{1L}^{ki}}^\ast{N_{2 R}^{ki}} \mbox{F}_2(x^N_{ik})\right) \\
a^N_r & = & \sum_{i=1}^4 \sum_{k=1}^6 \frac{m^2_{\mu}}{8 \pi^2
m^2_{\tilde{l}_k}}\left( {N_{1R}^{ki}}^\ast{N_{2 R}^{ki}}
\mbox{F}_1(x^N_{ik}) +
\frac{m_{\tilde{\chi}^{0}_i}}{m_{\mu}}{N_{1R}^{ki}}^\ast{N_{2
L}^{ki}} \mbox{F}_2(x^N_{ik}) \right) \eeqa where
$x^N_{ik}=m^2_{\tilde{\chi}^{0}_i}/m^2_{\tilde{l}_k}$ and
\beqa\label{eq505}
\mbox{F}_1(x) & = & \frac{2x^3 + 3x^2 - 6x + 1 -6x\ln x}{12(x-1)^4} \nonumber \\
\mbox{F}_2(x) & = & \frac{x^2 - 1 -2x\ln x}{2(x-1)^3} \nonumber \\
N_{lL}^{ki} & = & \sqrt{2} g_{1}X_{i1}D^\ast_{l+3,k} + X_{i3}D^\ast_{lk}Y^{e(D)}_l\nonumber \\
N_{lR}^{ki} & = &
-\frac{1}{\sqrt{2}}(g_{1}X_{i1}+g_{2}X_{i2})D^\ast_{lk}+X_{i3}D^\ast_{l+3,k}Y^{e(D)}_l
\eeqa where $g_1$ and $g_2$ are the U(1) and SU(2) gauge coupling
constants, $X$ is the matrix diagonalizing the $4\times4$
neutralino mass matrix $M_{\tilde{\chi}^0}$ according to
$M_{\tilde{\chi}^0}X={X}^{\textstyle{\ast}}M_{\tilde{\chi}^0}^{(D)}$,
$D$ diagonalizes the $6\times6$ charged slepton mass matrix
$M_{\tilde{l}}^2$ according to
$M_{\tilde{l}}^{2}D=DM_{\tilde{l}}^{2(D)}$ and $Y^{e(D)}$ is the
diagonalized Yukawa matrix of charged leptons (We use the notation
of \cite{ADS}).

Similarly, the chargino diagram gives \beqa\label{eq506}
a^C_l & = & -\sum_{i=1}^2 \sum_{k=1}^6 \sum_{m=1}^3 \frac{m^2_{\mu}}{8 \pi^2 m^2_{\tilde{\nu}_k}}
            \left( {C_{1L}^{ki}}^\ast{C_{2 L}^{ki}} \mbox{F}_3(x^C_{ik}) +
            \frac{m_{\tilde{\chi}^{\pm}_i}}{m_{\mu}}{C_{1L}^{ki}}^\ast{C_{2 R}^{ki}} \mbox{F}_4(x^C_{ik})\right) \\
a^C_r & = & -\sum_{i=1}^2 \sum_{k=1}^6 \sum_{m=1}^3
\frac{m^2_{\mu}}{8 \pi^2 m^2_{\tilde{\nu}_k}}\left(
{C_{1R}^{ki}}^\ast{C_{2 R}^{ki}} \mbox{F}_3(x^C_{ik}) +
\frac{m_{\tilde{\chi}^{\pm}_i}}{m_{\mu}}{C_{1R}^{ki}}^\ast{C_{2
L}^{ki}} \mbox{F}_4(x^C_{ik}) \right) \eeqa where
$x^C_{ik}=m^2_{\tilde{\chi}^{\pm}_i}/m^2_{\tilde{\nu}_k}$ and
\beqa\label{eq507}
\mbox{F}_3(x) & = & \frac{x^3 - 6x^2 + 3x + 2 + 6x\ln x}{12(x-1)^4} \nonumber \\
\mbox{F}_4(x) & = & \frac{x^2 - 4x + 3 + 2\ln x}{2(x-1)^3} \nonumber \\
C_{lL}^{ki} & = & -g_{2} V_{i1} P_{mk}^{\ast} (V_{MNS})^\ast_{lm} + V_{i2} P_{m+3,k}^{\ast}
                Y^{\nu (D)}_m (V_{MNS})^\ast_{lm} \nonumber \\
C_{lR}^{ki} & = & U_{i2}^{\ast} P_{mk}^{\ast} (V_{MNS})^\ast_{lm}
Y^{e(D)}_l \eeqa where $V_{MNS}$ is the MNS mixing matrix, $P$ is
the matrix diagonalizing the $6\times6$ sneutrino mass matrix
$M_{\tilde{\nu}}^2$ according to
$M_{\tilde{\nu}}^{2}P=PM_{\tilde{\nu}}^{2(D)}$, $U$ and $V$
diagonalize the chargino mass matrix $M_{\tilde{\chi}^{\pm}}$
according to $U^{\textstyle{\ast}} M_{\tilde{\chi}^{\pm}} V^{\dag}
= M_{\tilde{\chi}^{\pm}}^{(D)}$ and $Y^{\nu (D)}$ is the
diagonalized Yukawa matrix of neutrinos.

To evaluate the branching ratio of $\muegamma$, we first generate
Yukawa textures in the way described in Sec.~3 with
phenomenological inputs including the fermion masses and neutrino
oscillations described in Sec.~4. Then we choose soft breaking
parameters at the GUT scale and run the RGEs to the weak scale.
Finally one can use the formula given above to calculate the
$\muegamma$ branching ratio. We display our results in the
following three figures for $\tan\beta = 10$, 30 and 40.

\begin{figure}[h]
\begin{center}
\scalebox{0.6}[0.6]{\includegraphics{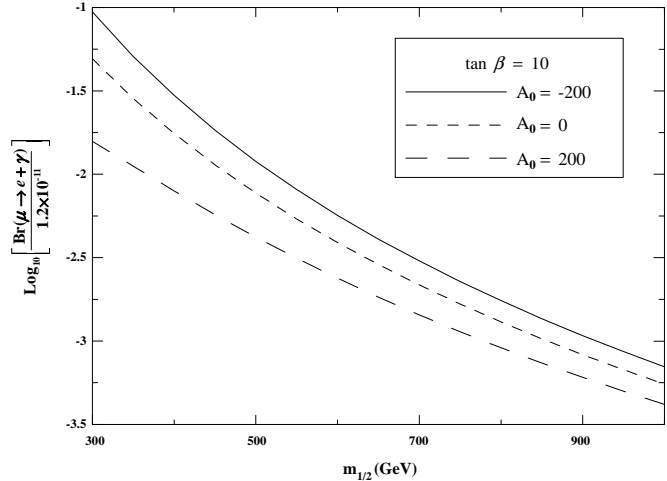}}
\end{center}
\caption{\label{10} Branching ratio of $\mu\rightarrow e+ \gamma$
for $\tan\beta=10$}
\end{figure}

\begin{figure}[h]
\begin{center}
\scalebox{0.6}[0.6]{\includegraphics{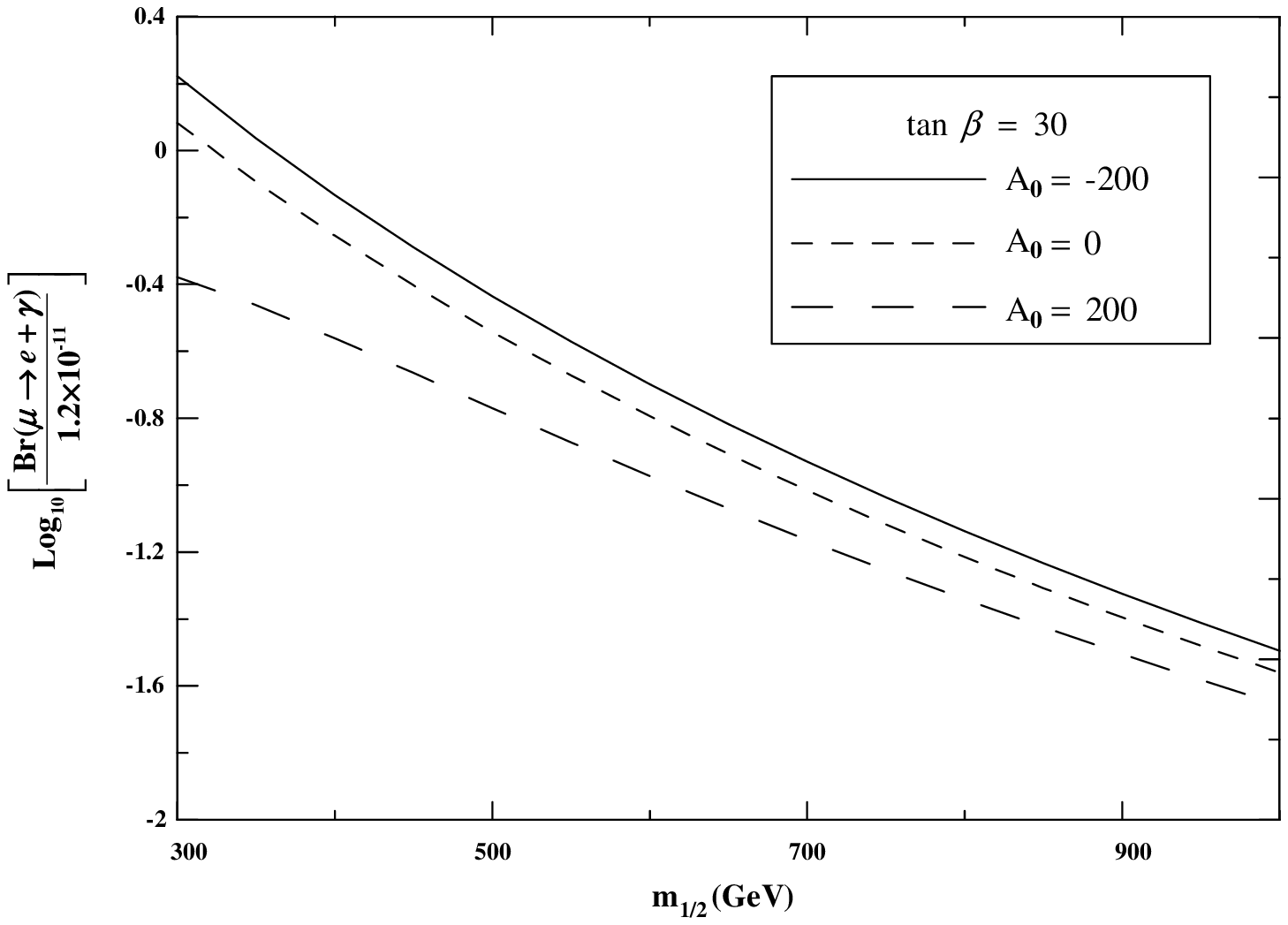}}
\end{center}
\caption{\label{30} Branching ratio of $\mu\rightarrow e+ \gamma$
for $\tan\beta=30$}
\end{figure}

\begin{figure}[h]
\begin{center}
\scalebox{0.60}[0.60]{\includegraphics{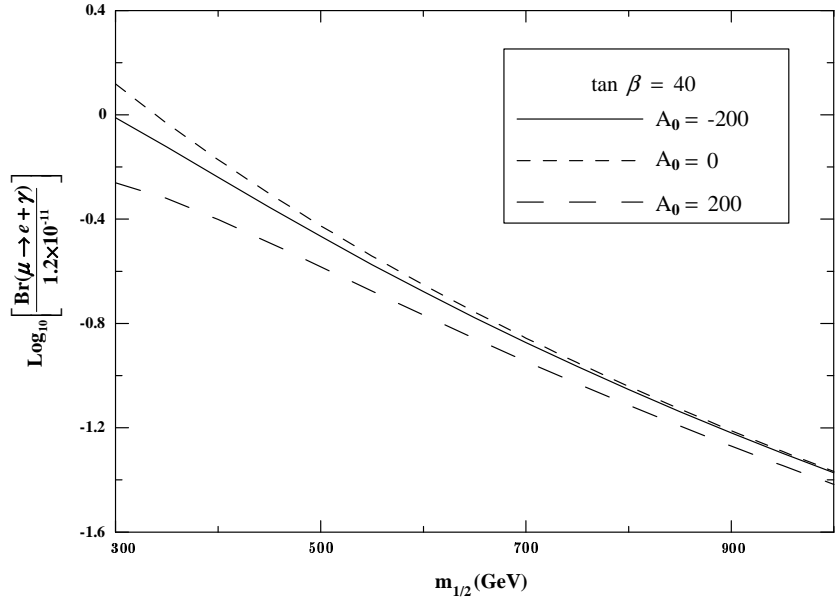}}
\end{center}
\caption{\label{40} Branching ratio of $\mu\rightarrow e+ \gamma$
for $\tan\beta=40$}
\end{figure}

Although our Yukawa textures are constructed through the Kahler
potential, the usual mSUGRA structure holds at the GUT scale. In
addition, to constrain the parameter space, we implement the relic
density constraint \cite{wmap}: $0.095\leq \Omega_{\chi^0_1}h^2
\leq 0.129$. For a given $m_{1/2}$, an allowed narrow  region of
$m_0$ is determined by the relic density constraint \cite{ADS,
cnst}. Since the $\mu\rightarrow e+ \gamma$ branching ratio is not
sensitive to the value of $m_0$ we only show the result for one
value of $m_0$ in the allowed region for any given $m_{1/2}$.
There are other experimental bounds on the parameter space, e.g.
the $b\rightarrow s + \gamma$ decay, the light Higgs mass, the muon $g-2$, all of which
can easily be  implemented. Since $m_0$ is significantly
constrained by the relic density bounds for any given $m_{1/2}$,
other experimental constraints are only needed for constraining
$m_{1/2}$. For example, at $\tan\beta = 40$, the $b\rightarrow s +
\gamma$ branching ratio produces a lower bound of $\simeq 400
\;\mbox{GeV}$ on $m_{1/2}$ while a muon $g-2$ deviation from the
SM can produce an upper bound on $m_{1/2}$. For the purpose of
showing what $\muegamma$ branching ratio can be reached in our
model, except for the relic density constraint, we ignore the
other experimental bounds in our plots since they are not
significant for this purpose.

In our plots the y-axis is the logarithmic ratio of our
theoretical predictions to the current experimental bound
\cite{PDG02a}. Therefore, only the region below zero is
experimentally allowed. One can see that at large $\tan\beta$,
especially for $\tan\beta=40$, the theory predictions are only
about one order of magnitude smaller than the experimental bound
and hence accessible to future experiments \cite{Mori99, MECO}
while part of the parameter space for lower $\tan\beta$ will also
be accessible.

\section{conclusion}

In this paper we have extended a model of the quark mass hierarchy
based on the \mbox{Horava-Witten} M-Theory \cite{ab} to include charged
leptons and massive neutrinos. The model is based on the
assumptions that five branes exists in the bulk lying near the
distant orbifold plane (i. e. about 90\% of the way from the
physical plane), and that the instanton charges on the physical
plane vanish. We had previously seen that this gave rise to a
three generation model with the Standard Model gauge group at the
GUT scale. While one cannot calculate Yukawa couplings in M-Theory
(they involve integrals over the Calabi-Yau space) these
constraints were sufficient to qualitatively account for the quark
mass hierarchy at the electroweak scale without undue fine tuning.
The mechanism that achieved this was that the five brane
contribution to the Kahler potential gave rise to small Kahler
matrix eigenvalues, and the quark masses were proportional to the
reciprocal square root of the eigenvalues when the kinetic energy
was put into canonical form. We saw that the same mechanism also
gave rise qualitatively to the hierarchy of charged lepton masses,
again without any excessive fine tuning.

Neutrino masses can arise in these models if a right handed
neutrino exists in the massless particle spectrum. Then one can
assume that the Kahler potential has a cubic holomorphic
contribution of the form of Eq.(\ref{eq402}), the interaction
being scaled by the 11 dimensional Planck mass (the basic
parameter of Horava-Witten theory). When transformed to the
superpotential by a Kahler transformation, this term gives rise to
neutrino masses of the correct size after supersymmetry breaking.
(Thus the mechanism being used here for the neutrino masses is
similar to the one previously used to generate a $\mu$ parameter
of electroweak size \cite{muterm}.) it is possible then to chose
natural sized values for the Yukawa and Kahler matrix entries to
generate masses and CKM and MNS mixing angles in agreement with
all low energy data.  The neutrinos in this model are Dirac, and
so will exclude neutrinoless double beta decay. However, the
mixing in the neutrino sector allows for $\muegamma$ decay to
occur, and with reasonable values of the SUSY parameters, this
decay should become observable in the next round of $\mu$
catalysis experiments \cite{Mori99,MECO} over a significant range
of parameters.

Aside from the Kahler and Yukawa matrices, the quark, lepton and
neutrino properties depend on the Calabi-Yau volume modulus
$\mathcal{V}$ which we have parameterized by $V^{1/6}=r$ and the
radius modulus R. We have found that all the quark, lepton and
neutrino masses can be fit satisfactorily with $r$ and $R$ of
$O(1)$. Thus for the example in text for $\tan\beta=40$ we found
$R=2.13$ and $r=3.20$. One important feature of this Horava-Witten
model that has not been addressed here is how to stabilize the
position of the 5-brane close to the distant orbifold plane. One
possibility may involve quantum corrections, e.g. membrane
potentials between the 5-brane and the orbifold planes \cite{MPS,
Lima, Curio}.

\begin{acknowledgments}
This work was supported in part by National Science Foundation
Grant PHY-0101015 and in part by the Natural Science and
Engineering Research Council of Canada.
\end{acknowledgments}

\end{document}